\documentclass[prl,aps,floats,twocolumn]{revtex4}

\usepackage{graphicx}
\usepackage{amssymb}
\usepackage{amsmath}
\usepackage{float}
\usepackage{color}

\begin{document}

\title{Reflection Quasilattices and the Maximal Quasilattice}

\author{Latham Boyle$^1$ and Paul J. Steinhardt$^{1,2}$}

\affiliation{$^{1}$Perimeter Institute for Theoretical Physics, Waterloo, Ontario N2L 2Y5, Canada \\
$^{2}$ Princeton Center for Theoretical Science and Department of Physics, \\
Princeton University, Princeton, New Jersey 08544, USA}
  
\date{April 2016}
                            
\begin{abstract}
We introduce the concept of a {\it reflection quasilattice}, the quasiperiodic generalization of a Bravais lattice with irreducible reflection symmetry.   Among their applications, reflection quasilattices are the reciprocal ({\it i.e.}\ Bragg diffraction) lattices for quasicrystals and quasicrystal tilings, 
such as Penrose tilings, with irreducible reflection symmetry and discrete scale invariance.  In a follow-up paper, we will show that reflection quasilattices can be used to generate tilings in real space with properties analogous to those in Penrose tilings, but with different symmetries and in various dimensions. 
Here we explain that reflection quasilattices only exist in dimensions two, three and four, and we prove that there is a unique reflection quasilattice in dimension four: the ``maximal reflection quasilattice" in terms of dimensionality and symmetry.  Unlike crystallographic Bravais lattices, all reflection quasilattices are invariant under rescaling by certain discrete scale factors.  We tabulate the complete set of scale factors for all reflection quasilattices in dimension $d>2$, and for all those with quadratic irrational scale factors in $d=2$. 
\end{abstract}

\maketitle 

\section{Introduction}  

Our starting point is Coxeter's celebrated classification of the finite reflection groups in terms of irreducible root systems and Coxeter-Dynkin diagrams \cite{Cox1, Cox2, Cox3, RegularPolytopes, CoxeterMoser}.  These come in two flavors: crystallographic and non-crystallographic (see Refs.~\cite{ConwaySloane, Humphreys1}).  We introduce two simple definitions: if $\Phi$ is an irreducible (crystallographic or non-crystallographic) root system, and $G(\Phi)$ is the corresponding reflection group, the $\Phi$ root (quasi)lattice $\Lambda_{\Phi}$ is the set of all integer linear combinations of the $\Phi$ roots; and a $\Phi$ reflection (quasi)lattice is any $G(\Phi)$-invariant subset of $\Lambda_{\Phi}$ that is closed under addition and subtraction.  (The prefix ``quasi," referring to quasiperiodic, is used when $\Phi$ is non-crystallographic.)  Thus, $\Lambda_{\Phi}$ is always one of the $\Phi$ reflection (quasi)lattices.  

In this paper, we explain why the reflection quasilattices (``reflection QLs") are of particular interest, and why the above definitions are particularly apt.  We show that the $d$-dimensional reflection QLs are the non-crystallographic generalization of the special class of $d$-dimensional Bravais lattices whose point symmetry is an irreducible reflection group $G(\Phi)$ of full rank $d$.  They play a fundamental physical role: first, as the reciprocal ({\it i.e.}\ Bragg diffraction) lattices for quasicrystals \cite{LevineSteinhardt84, Janot, Senechal, Baake2002, SteurerDeloudi, BaakeGrimme} and quasicrystal patterns/tilings \cite{GrunbaumShephard} with point symmetry $G(\Phi)$; and second, as the basis for classifying the space groups corresponding to $G(\Phi)$ \cite{Mermin88, Mermin92, BoyleSteinhardt2}.  Reflection QLs have two other key properties of physical interest.  (i) First, their discrete point symmetry $G(\Phi)$ is sufficient to rigidly fix their shape (with no continuously tunable parameters, apart from overall rescaling); and, consequently, the associated quasicrystals and tilings are governed by certain irrational ratios that are ``locked in" by symmetry, so that they do not require any fine tuning and are intrinsically robust ({\it e.g.}\ against fluctuations of temperature and pressure in the laboratory).  (ii) Second, in contrast to any ordinary lattice, and in contrast to the broader class of quasilattices defined in previous work \cite{Mermin1987a, Mermin1987b}, every reflection QL is precisely invariant under rescaling by a special set of characteristic {\it scale factors}, and thus exhibits a discrete scale invariance which suggests interesting connections to other instances of scale invariance in physics.   Reflection QLs are the reciprocal lattices for many of the most experimentally and/or mathematically interesting and widely studied quasicrystals and tilings, including: (i) the Penrose tiling \cite{Penrose74, Gardner77, Penrose78}; (ii) natural generalizations of it that share its key properties (including its Ammann-grid decoration \cite{SocolarSteinhardt86, Socolar89, BoyleSteinhardt1, BoyleSteinhardt2}); and (iii) a variety of other quasicrystals obtained from higher-dimensional root lattices by the cut-and-project method \cite{BaakeRootLattices, ElserSloane, SadocMosseri1, SadocMosseri2, MoodyPatera, BaakeGahler, BaakeGrimme}.   In fact, in a forthcoming paper \cite{BoyleSteinhardt2}, we show how reflection QLs can be used to systematically generate tilings with matching rules, inflation rules, and Ammann-grid decorations analogous to those in Penrose tilings (but with different symmetries, and in various dimensions), in a way that illuminates the deep web of connections between aperiodic translational order, non-crystallographic orientational order, and discrete scale invariance that these tilings embody.

In this paper, we explain that reflection QLs only occur in two, three and four dimensions, and prove that there is a unique reflection QL in four dimensions (4D), which we present explicitly.  This is the final and maximal reflection QL -- the reflection QL of highest dimension and highest symmetry.  We discuss the relation of these results to earlier work \cite{Mermin1987a, Mermin1987b} on a related class of quasilattices in two dimensions (2D) and three dimensions (3D), and to number theoretic results about the quaternions \cite{Reiner, Tits, Vigneras, MoodyWeiss, BaakeGrimme}.  

For all the reflection QLs in dimension $d>2$, and all the {\it quadratic} reflection QLs ({\it i.e.}\ reflection QLs with quadratic irrational scale factors) in dimension $d=2$, we tabulate the corresponding scale factors, and prove that we have the complete set.

\section{Non-crystallographic root systems}  

For an introduction to finite reflection groups (finite Coxeter groups), roots systems and Coxeter-Dynkin diagrams, see Chapter 4, Section 2 in \cite{ConwaySloane} (for a brief introduction) and Part 1 ({\it i.e.}\ Chs. 1-4) in \cite{Humphreys1} (for more detail).  For an introduction to a wide range of relevant mathematics underlying our study, see \cite{BaakeGrimme}.

%\begin{figure}
%  \begin{center}
%    \includegraphics[width=2.8in]{diagrams.eps}
%  \end{center}
%  \caption{The Coxeter-Dynkin diagrams for the finite irreducible root systems (equivalently, the finite irreducible reflection groups).  The non-crystallographic %cases are boxed.}
% \label{diagrams}
%\end{figure}

The irreducible finite reflection groups and their corresponding root systems may be neatly described by Coxeter-Dynkin diagrams (see \cite{ConwaySloane, Humphreys1}).  These come in two varieties: crystallographic and non-crystallographic.  The crystallographic cases are familiar from the theory of Lie groups and Lie algebras: they come in four infinite families ($A_{n}$, $B_{n}$, $C_{n}$ and $D_{n}$) and five exceptional cases ($G_{2}$, $F_{4}$, $E_{6}$, $E_{7}$ and $E_{8}$).  The remaining roots systems are non-crystallographic: almost all of these are in 2D ($I_{2}^{n}$, $n=5,7,8,9,\ldots$), with just one in 3D ($H_{3}$), one in 4D ($H_{4}$), and none in higher dimensions.

Let us describe the non-crystallographic roots systems:

First consider $I_{2}^{n}$.  In geometric terms, the $2n$ roots of $I_{2}^{n}$ are perpendicular to the $n$ mirror planes of an equilateral $n$-sided polygon; note that when $n$ is odd, these mirror planes are all equivalent (each intersects a vertex and its opposite edge), but when $n$ is even the mirror planes split into two inequivalent sets (those that intersect two opposite vertices, and those that intersect two opposite edges).  In algebraic terms, we can think of the $2n$ roots as $2n$ complex numbers.  When $n$ is odd, these are the $(2n)$th roots of unity: $\zeta_{2n}^{k}$ ($k=1,\ldots,2n$), where $\zeta_{n}\equiv{\rm exp}(2\pi i/n)$.  When $n$ is even, the $2n$ roots break into two rings: (i) a first ring $\zeta_{n}^{k}$ ($k=1,\ldots,n$); and (ii) a second ring which we can think of as $\zeta_{n}^{k}+\zeta_{n}^{k+1}$ ($k=1,\ldots,n$).  The $I_{2}^{n}$ reflections generate the symmetry group of the regular $n$-gon, of order $2n$.

Next consider $H_{3}$.  If we let $\tau$ denote the golden ratio $\frac{1}{2}(1+\sqrt{5})$, the $H_{3}$ roots are the 30 vectors obtained from
\begin{equation}
  \label{H3_roots}
    \{\pm1,0,0\}\quad{\rm and}\quad\frac{1}{2}\{\pm\tau,\pm1,\pm1/\tau\}
\end{equation}
by taking all combinations of $\pm$ signs, and all even permutations of the three coordinates.  These point to the 30 edge midpoints of a regular icosahedron, and the corresponding reflections generate the full symmetry group of the icosahedron (of order 120).

Finally consider $H_{4}$.  From a geometric standpoint, the $H_{4}$ roots are the 120 vectors obtained from
\begin{equation}
  \label{H4_roots}
  \begin{array}{c}
    \{\pm1,0,0,0\} \\
    (1/2)\{\pm1,\pm1,\pm1,\pm1\} \\
    (1/2)\{\;\!\!0,\pm\tau,\pm1,\pm1/\tau\;\!\!\}
  \end{array}
\end{equation}
by taking all combinations of $\pm$ signs, and all even permutations of the four coordinates: these are the 120 vertices of a 4D regular polytope called the 600 cell \cite{RegularPolytopes}.  From the algebraic standpoint, they are the set of 120 ``unit icosians" \cite{ConwaySloane} within the skew field of quaternions $\mathbb{H}$ (see \cite{ConwaySmith} for an introduction).  The $H_{4}$ reflections generate the symmetry group of the 600 cell: this group has $120^{2}=14400$ elements, corresponding to all maps from $\mathbb{H}\to\mathbb{H}$ of the form $Q\to\bar{q}_{1}Q q_{2}$ or $Q\to\bar{q}_{1}\bar{Q}q_{2}$, where $q_{1}$ and $q_{2}$ are any two unit icosians \cite{ConwaySmith, ElserSloane}.  

\section{Root and reflection lattices and quasilattices: definitions} 

Let $\Phi$ be a finite irreducible (crystallographic or non-crystallographic) root system, with 
$G(\Phi)$ the corresponding reflection group.  We introduce two  definitions:

{\it Definition 1.} The $\Phi$ ``root (quasi)lattice" $\Lambda_{\Phi}$ is the set of all integer linear combinations of the $\Phi$ roots. 

{\it Definition 2.} A $\Phi$ ``reflection (quasi)lattice" is any subset of $\Lambda_{\Phi}$ (including $\Lambda_{\Phi}$ itself) that is: (i) $G(\Phi)$-invariant and (ii) closed under addition and subtraction.

Here ``quasi" is used when $\Phi$ is non-crystallographic, and we abbreviate quasilattice as QL.  
The term ``lattice" without any prefix refers to crystallographic only.

Let $\Phi$ have rank $d$: if $\Phi$ is crystallographic, a $\Phi$ root (or reflection) lattice is an ordinary lattice in $\mathbb{R}^{d}$ (with some finite minimum separation between nearest neighbors); while if $\Phi$ is non-crystallographic, then a $\Phi$ root (or reflection) QL is a {\it dense} set of points in $\mathbb{R}^{d}$ (with points arbitrarily close to every point in $\mathbb{R}^{d}$).

\section{Remarks on these definitions}  

Definition 1 is clear: it is the non-crystallographic generalization of a (crystallographic) root lattice.  But to fully appreciate Definition 2, it is helpful to review the definition of a ``$G$-lattice" proposed by Rokhsar, Mermin and Wright (RMW) \cite{Mermin1987b}:  

Let $G$ be a point group in $\mathbb{R}^{d}$, and let $\Lambda$ be a rank-$d$ set of vectors in $\mathbb{R}^{d}$. $\Lambda$ is a $d$-dimensional $G$-lattice if it: (i) is $G$-invariant; (ii) is closed under addition and subtraction; and (iii) is of the minimal integer rank compatible with $G$-invariance.  This is the non-crystallographic generalization of the idea of a Bravais lattice with point group $G$ (since, when $G$ is crystallographic, the $G$-lattices are precisely the Bravais lattices with point group $G$).  

{\it Remark 1.} {\it The reflection (quasi)lattices are a natural subclass of Bravais (quasi)lattices: those whose point group is an irreducible reflection group of full rank.}  That is, the reflection (quasi)lattices are the $d$-dimensional $G$-lattices for which $G=G(\Phi)$ is an irreducible rank $d$ reflection group.  This can be proved as follows.  First of all, one can check that any set of vectors that is $G(\Phi)$-invariant and closed under addition and subtraction must contain a copy of the $\Phi$ root system itself.  [Let us check the $H_{4}$ case to illustrate: if $\Lambda$ is $H_{4}$-symmetric and $\lambda=\{w,x,y,z\}$ is any element in $\Lambda$ with $w\neq0$, then by an $H_{4}$ transformation, $\lambda'=\{w,-x,-y,-z\}$ is also in $\Lambda$, and hence so is $\lambda+\lambda'=\{2w,0,0,0\}$, which is $2w$ times the $H_{4}$ root $\{1,0,0,0\}$.  Thus, by $H_{4}$-symmetry, $\Lambda$ must contain the whole $H_{4}$ root system (times $2w$).]  We draw two implications from this.  First, a $\Phi$ reflection (quasi)lattice has the minimal integer rank compatible with $G(\Phi)$ invariance, and is hence a $G(\Phi)$-lattice. Second, by the ``geometric lemma" in \cite{Mermin1987b} (which says that, if $\Lambda$ has integer rank $n$, then any $n$ integrally independent vectors in $\Lambda$ integrally span $\Lambda$, after a suitable rescaling) it follows that any $G(\Phi)$-lattice is integrally spanned by the $\Phi$ roots, and is hence a $\Phi$ reflection (quasi)lattice.  

{\it Remark 2.} {\it Compared to the earlier approach of defining and studying the class of $G$-lattices, our approach of defining and studying the class of reflection lattices and quasilattices has key advantages.}  (i) Speaking first in general terms, Definition 2 is a more elegant starting point than the definition of a $G$-lattice, and more connected to the heart of mathematics via root lattices.  (ii) In particular, the $G$-lattice definition relies on the awkward minimal-integer-rank condition, which is needed to exclude a host of other, less interesting, ``non-minimal" or ``incommensurately modulated" crystals  and quasicrystals \cite{Mermin92}.  By contrast, Definition 2 has the conceptual advantage that, since it is fundamentally based on the notion of a root system, the minimal-integer-rank property is achieved  {\it automatically}, with no need to impose this as a separate condition.  (iii) Similarly, the most interesting quasicrystals and aperiodic tilings (such as the Penrose tiling) exhibit discrete scale invariance: reflection quasilattices have this property automatically, whereas $G$-quasilattices do not.  (iv) From Definition 2 and Remark 1, we infer another feature that is interesting, both mathematically and physically: a reflection (quasi)lattice is a special type of Bravais (quasi)lattice whose shape is completely ``locked in" by its point group (with no tunable shape parameters, apart from overall rescaling).

\section{The reflection quasilattices}  

{\it Reflection QLs in two dimensions:}  For each integer $n=5,7,8,9\ldots$, the $I_{2}^{n}$ root QL is the ring of cyclotomic integers $\mathbb{Z}(\zeta_{n})$ -- {\it i.e.}\ the set of all integer linear combinations of the $n$th roots of unity $\zeta_{n}^{k}$.  (Note: when $n$ is odd, the $I_{2}^{n}$ and $I_{2}^{2n}$ root QLs are redundant.)  By the logic in \cite{Mermin1987b} (see also  \cite{WashingtonCyclotomic, Edwards}), all {\it other} $I_{2}^{n}$ reflection QLs correspond to non-trivial ideals within $\mathbb{Z}(\zeta_{n})$: finding all such ideals is an important unsolved problem in algebraic number theory, so we cannot enumerate all the $I_{2}^{n}$ reflection QLs for general $n$; but for all $n<23$, and all even $n<46$, the only $I_{2}^{n}$ reflection QL is the $I_{2}^{n}$ root QL.

{\it Reflection QLs in three dimensions:}  The argument in Sec.~3 of \cite{Mermin1987b} implies that there are precisely three reflection QLs in 3D (all of type $H_{3}$).  To describe them, first recall that the $H_{3}$ roots (\ref{H3_roots}) point to the edge midpoints of a regular icosahedron.  The 12 vertices of this icosahedron are then the 12 vectors obtained from $\{\pm1,\pm\tau,0\}$ by taking all combinations of $\pm$ signs and all even permutations of the coordinates.  Now choose $v_{1},\ldots,v_{6}$ to be six of these vectors that are integrally independent ({\it e.g.}\ the six vectors obtained from $\{1,\pm\tau,0\}$ by including both $\pm$ options, and all even permutations of the coordinates).  The three reflection QLs ($H_{3}^{1}$, $H_{3}^{2}$ and $H_{3}^{3}$) consist of all linear combinations $m_{1}v_{1}+\ldots+m_{6}v_{6}$, where the coefficients satisfy an appropriate restriction: for $H_{3}^{1}$, the coefficients $m_{i}$ must be integers; for $H_{3}^{2}$, the coefficients $m_{i}$ must be integers whose sum $m_{1}+\ldots+m_{6}$ is even; and for $H_{3}^{3}$, the coefficients $m_{i}$ must either be all integers ($m_{i}\in\mathbb{Z}$) or all half-integers ($m_{i}\in\mathbb{Z}+\frac{1}{2}$).  We refer to these three $H_{3}$ reflection QLs as ``primitive", ``fcc" and ``bcc", respectively, since they arise by orthogonally projecting the six-dimensional primitive cubic, fcc or bcc lattices, respectively, on a maximally-symmetric 3D subspace.

{\it The maximal reflection QL:}   We next prove that there is a unique reflection QL in 4D.  This is the ``maximal reflection QL," maximal in terms of both dimensionality and symmetry.  

First note that the only available root system for a reflection QL $\Lambda$ in 4D is $H_{4}$.  Every vector $\lambda\in\Lambda$ can then be written as an integer linear combination of the 120 $H_{4}$ roots, which can, in turn, be written as an integer linear combination of the eight vectors
\begin{equation}
  \begin{array}{llll} 
  \{\frac{1}{2},0,0,0\},&\{0,\frac{1}{2},0,0\},&\{0,0,\frac{1}{2},0\},&\{0,0,0,\frac{1}{2}\}, \\
  \{\frac{\tau}{2},0,0,0\},&\{0,\frac{\tau}{2},0,0\},&\{0,0,\frac{\tau}{2},0\},&\{0,0,0,\frac{\tau}{2}\}.
  \end{array}
\end{equation}
So the subset of vectors in $\Lambda$ that are proportional to $\{1,0,0,0\}$ can all be written in the form
$(1/2)\{m+n\tau,0,0,0\}$ (with $m,n\in\mathbb{Z}$) -- {\it i.e.}\ they are a subset of (a scaled copy of) the ``golden integers" (the set of numbers $m+n\tau$ with $m,n\in\mathbb{Z}$).  Now consider any such vector $\{w,0,0,0\}$.  By $H_{4}$ symmetry, $\Lambda$ must also also contain $w$ times every $H_{4}$ root and, in particular, it must contain $(w/2)\{\tau,1/\tau,1,0\}$ and $(w/2)\{\tau,-1/\tau,-1,0\}$ as well as their sum $\tau\{w,0,0,0\}$.  Hence we can apply the ``algebraic lemma" proved in \cite{Mermin1987a} (which says that any subset of the golden integers that is closed under addition and subtraction and scaling by $\tau$ must be a scaled copy of the golden integers) to infer that the subset of vectors in $\Lambda$ that are proportional to $\{1,0,0,0\}$ are a scaled copy of the golden integers.  Let us rescale the QL so that the vectors proportional to $\{1,0,0,0\}$ are precisely the set of all golden integers times $\{1,0,0,0\}$ (and, by symmetry, the vectors proportional to any root are precisely the set of all golden integers times that root).  So $\Lambda$ {\it must} contain all the golden integers times each $H_{4}$ root, and all integer linear combinations of such vectors.  We will next show that it cannot contain anything else.

To see this, first recall that if $\Lambda$ contains a vector $\lambda=\{w,x,y,z\}$, it also contains the vector $\{2w,0,0,0\}$; and, by a similar argument, it also contains the vectors $\{0,2x,0,0\}$, $\{0,0,2y,0\}$ and $\{0,0,0,2z\}$.  Since $2w$, $2x$, $2y$ and $2z$ must all be golden integers, it follows that any vector $\lambda\in\Lambda$ must have the form
\begin{equation}
  \label{lambda_H4}
  \lambda=(1/2)\{m_{0}+n_{0}\tau,m_{1}+n_{1}\tau,m_{2}+n_{2}\tau,m_{3}+n_{3}\tau\}
\end{equation}
(with $m_{i},n_{i}\in\mathbb{Z}$).  But if we apply any $H_{4}$ transformation to $\lambda$, the requirement that the new vector $\lambda'$ must also have this form (with new integers $m_{i}',n_{i}'$) restricts the possible values of the integers $m_{i}$ and $n_{i}$.  It is enough to consider the transformation $\lambda'=\lambda q$ where $q$ is any unit icosian (\ref{H4_roots}); in this way we obtain the constraints
\begin{equation}
  \label{constraints}
  \begin{array}{rcl}
    m_{\alpha}+m_{\beta}+m_{\gamma}+m_{\delta}&=&{\rm even}  \\
    n_{\alpha}+n_{\beta}+n_{\gamma}+n_{\delta}&=&{\rm even} \\ 
    m_{\alpha}+n_{\alpha}+m_{\beta}+n_{\gamma}&=&{\rm even}, 
  \end{array}
\end{equation}
where the indices $\{\alpha,\beta,\gamma,\delta\}$ are any even permutation of $\{0,1,2,3\}$.  In considering which combinations of $m_{i}$ and $n_{i}$ are allowed, it is also enough to consider $m_{i}$ and $n_{i}$ to be valued mod 2, since we already know that $\Lambda$ contains all golden integers times the four cartesian unit vectors, so if it contains the combination $\{m_{0},m_{1},m_{2},m_{3},n_{0},n_{1},n_{2},n_{3}\}$, it also contains the combination where one or more of these integers is shifted by $\pm2$.  Thus, we can simply enumerate all 16 allowed vectors (\ref{lambda_H4}) satisfying the constraints (\ref{constraints}): namely, $\{0,0,0,0\}$, $\frac{1}{2}\{1,1,1,1\}$, $\frac{1}{2}\{\tau,\tau,\tau,\tau\}$, $\frac{1}{2}\{1+\tau,1+\tau,1+\tau,1+\tau\}$ and all even permutations of $\frac{1}{2}\{0,1+\tau,\tau,1\}$.  But each of these vectors is a golden integer times a root, which we already proved {\it had} to be in $\Lambda$.

Thus, an $H_{4}$ reflection QL must contain all integer linear combinations of the $H_{4}$ roots, and nothing else -- this completes the proof that it is unique (and is none other than the $H_{4}$ root QL).  This corresponds to the ring of quaternions known as the icosians, which may be obtained by orthogonally projecting the $E_{8}$ root lattice on a maximally-symmetric 4D subspace \cite{ConwaySloane, ElserSloane}.  In fact, our geometric proof turns out to be ultimately equivalent to the number-theoretic result that every left ideal in the icosians is principal \cite{Reiner, Tits, Vigneras, MoodyWeiss, BaakeGrimme}.

\section{Discrete scale invariance}  

Unlike an ordinary lattice, which has no scale invariance, each reflection QL has discrete scale invariance -- it is exactly invariant under rescaling by any integer power of one or more ``scale factors."   Just as we cannot enumerate all the reflection QLs in 2D, we cannot enumerate all of their scale factors in 2D -- but we {\it can} say that the scale factors for the $I_{2}^{n}$ reflection QLs will be irrationals of order $\phi(n)/2$ \cite{WashingtonCyclotomic}, where Euler's totient function $\phi(n)$ is the number of positive integers less than $n$ (including 1) that share no common factor with $n$; and also that a {\it subset} of the scale factors for the $I_{2}^{n}$ {\it root} QL will be given by elementary expressions called the real cyclotomic units \cite{WashingtonCyclotomic}.

An important set of 2D reflection QLs are the three cases with $\phi(n)/2=2$: $I_{2}^{5}/I_{2}^{10}$, $I_{2}^{8}$ and $I_{2}^{12}$.  For these three 2D reflection QLs, and for the 3D  $H_3$ and 4D $H_4$ reflection QLs, the scale factors are quadratic irrationals.  Following similar reasoning to that in \cite{Mermin88, Mermin92, BaakeGrimme}, the complete set of possible scale factors can be derived explicitly by the following argument.  If a QL is invariant under rescaling by $\eta$, then any 1D sublattice must also be invariant under the same rescaling.  In other words, the scaling group of the QL must be a subgroup of the scaling group of its 1D sublattice.  Each of the $I_{2}^{5}/I_{2}^{10}$, $I_{2}^{8}$, $I_{2}^{12}$, $H_{3}$ and $H_{4}$ reflection QLs contain a 1D sublattice corresponding to a ring of real quadratic integers $\mathbb{Z}(\sqrt{\kappa})$: $I_{2}^{5}/I_{2}^{10}$, $H_{3}$ and $H_{4}$ contain $\mathbb{Z}(\sqrt{5})$, $I_{2}^{8}$ contains $\mathbb{Z}(\sqrt{2})$ and $I_{2}^{12}$ contains $\mathbb{Z}(\sqrt{3})$.  But the scale factors of $\mathbb{Z}(\sqrt{\kappa})$ (where $\kappa$ is a positive square-free integer) are precisely $\pm u^{k}$ where $k$ is any integer and the ``fundamental unit" $u$ is given by $(a+b\sqrt{\Delta})/2$ where $a$ and $b$ are the smallest positive integer solutions of $a^{2}-\Delta b^{2}=\pm4$ and $\Delta$ is the discriminant of $\mathbb{Q}(\kappa)$ ($\Delta=\kappa$ if $\kappa=1$ mod $4$, and $\Delta=4\kappa$ if $\kappa=2$ or $3$ mod $4$) \cite{Neukirch}.  So, for $\mathbb{Z}(\sqrt{5})$, $\mathbb{Z}(\sqrt{2})$ and $\mathbb{Z}(\sqrt{3})$, the fundamental units are $\tau=\frac{1}{2}(1+\sqrt{5})$ (the golden ratio), $1+\sqrt{2}$ (the silver ratio) and $2+\sqrt{3}$, respectively.  We then check that the reflection QL is symmetric under the full scaling group $\pm u^{k}$ of its 1D sublattice, except for $H_{3}^{1}$ which is invariant under the subgroup $\pm(u^{3})^{k}$.  

Table \ref{reflection QLs} summarizes our results.

\begin{table}
\begin{center}
\begin{tabular}{l|l|l}
Reflection quasilattice & Description & Scale factor \\
\hline
$I_{2}^{5}/I_{2}^{10}$ & root & $\tau$ \\
$I_{2}^{8}$ & root & $1+\sqrt{2}$ \\
$I_{2}^{12}$ & root & $2+\sqrt{3}$ \\
\hline
$H_{3}^{1}$ & primitive & $\tau^{3}$ \\
$H_{3}^{2}$ & fcc (root) & $\tau$ \\
$H_{3}^{3}$ & bcc & $\tau$ \\
\hline
$H_{4}$ & root & $\tau$
\end{tabular}
\end{center}
\caption{All reflection quasilattices (``reflection QLs") in dimension $d>2$, and all {\it quadratic} reflection QLs ({\it i.e.}\ reflection QLs with scale factor given by a quadratic irrational) in $d=2$.  Here $\tau\equiv\frac{1}{2}(1+\sqrt{5})$ is the golden ratio.  The reflection QLs that are also root quasilattices are labeled ``root", while the descriptions ``primitive/fcc/bcc" are explained in the text.}
\label{reflection QLs}
\end{table} 

\section{Discussion}  

We conclude by mentioning a few directions for future work.  First, we note that the crystallographic root lattices are among the most important lattices in mathematics, playing a key role in the study of Lie algebras, Lie groups, representation theory, quivers, catastrophes, singularity theory and other contexts \cite{Humphreys2, Gabriel, Arnold, ADEintro, ConwaySloane}.  As generalizations, the root QLs and reflection QLs may lead to some extensions of these applications.   Second, as will be detailed in a forthcoming paper \cite{BoyleSteinhardt2}, there is a direct relationship between reflection QLs and tesselations that have the matching-rule, inflation-rule, and Ammann-grid properties of Penrose tilings. Given our proof here that there are only a handful of reflection QLs in dimension $d>2$, we can determine the complete set of irreducible Penrose-like tilings in $d>2$.  Finally, on a more speculative level, it is natural to notice that the maximal reflection QL -- a very distinctive and beautiful object -- exists in 4D, which is also the apparent dimension of spacetime: it is interesting to consider the connections  to fundamental physics, or to novel discretization schemes for Euclideanized 4D field theory.

\section{Acknowledgements} 

Research at Perimeter Institute is supported by the Government of Canada through Industry Canada and by the Province of Ontario through the Ministry of Research and Innovation. L.B. also acknowledges support from an NSERC Discovery Grant.


\begin{thebibliography}{99}

\bibitem{Cox1} H.S.M.~Coxeter, ``Groups whose fundamental regions are simplexes," J.\ London Math.\ Soc.\ {\bf 6}, 132 (1931).

\bibitem{Cox2} H.S.M.~Coxeter, ``Discrete groups generated by reflections," Ann.\ Math.\ {\bf 35}, 588 (1934).

\bibitem{Cox3} H.S.M.~Coxeter, ``Complete enumeration of finite groups of the form $R_{i}^{2}=(R_{i}R_{j})^{k_{ij}}=1$", J.\ London Math.\ Soc.\ {\bf 1}, 21 (1935).

\bibitem{RegularPolytopes} H.S.M.~Coxeter, {\it Regular Polytopes} (Methuen \& Co.\ Ltd., London, 1948).

\bibitem{CoxeterMoser} H.S.M.~Coxeter and W.O.J.~Moser, {\it Generators and Relations for Discrete Groups} (Springer-Verlag, Berlin, 1980).

\bibitem{ConwaySloane} J.H.~Conway and N.J.A.~Sloane, {\it Sphere Packings, Lattices and Groups} (Springer-Verlag, New York, 1993).

\bibitem{Humphreys1} J.E.~Humphreys, {\it Reflection Groups and Coxeter Groups} (Cambridge University Press, Cambridge, 1990).

\bibitem{LevineSteinhardt84} D.~Levine and P.J.~Steinhardt, ``Quasicrystals: A New Class of Ordered Structures," Phys.\ Rev.\ Lett.\ {\bf 53}, 2477 (1984).

\bibitem{Janot} C.~Janot, {\it Quasicrystals} (Oxford University Press, Oxford, 1994).

\bibitem{Senechal} M.~Senechal, {\it Quasicrystals and Geometry} (Cambridge University Press, Cambridge, 1995).

\bibitem{Baake2002} M.~Baake, "A guide to mathematical quasicrystals." Quasicrystals. Springer Berlin Heidelberg, 2002. 17-48. [arXiv:math-ph/9901014].

\bibitem{SteurerDeloudi} W.~Steurer and S.~Deloudi, {\it Crystallography of Quasicrystals: Concepts, Methods and Structures} (Springer-Verlag, Berlin, 2009).

\bibitem{BaakeGrimme} M.~Baake and U.~Grimm, {\it Aperiodic Order.  Volume 1: A Mathematical Invitation} (Cambridge University Press, Cambridge, 2013).

\bibitem{GrunbaumShephard} B.~Grunbaum and G.~C.~Shephard, {\it Tilings and Patterns} (W.H.Freeman and Company, New York, 1987).

\bibitem{Mermin88} D.S.~Rokhsar, D.C.~Wright and N.D.~Mermin, "The two-dimensional quasicrystallographic space groups with rotational symmetries less than 23-Fold," Acta Cryst., Sect.\ A: Found.\ Adv.\ {\bf 44}, 197 (1988).

\bibitem{Mermin92} Mermin, N. David. ``The space groups of icosahedral quasicrystals and cubic, orthorhombic, monoclinic, and triclinic crystals." Reviews of modern physics 64, no. 1 (1992): 3.

\bibitem{BoyleSteinhardt2} L.~Boyle and P.J.~Steinhardt, ``Coxeter Pairs, Ammann Patterns, and Penrose-like Tilings," arXiv:1608.08215.

\bibitem{Mermin1987a} Mermin, N. David, Daniel S. Rokhsar, and David C. Wright, ``Beware of 46-fold symmetry: The classification of two-dimensional quasicrystallographic lattices," Physical review letters 58.20 (1987): 2099.

\bibitem{Mermin1987b} Rokhsar, Daniel S., N. David Mermin, and David C. Wright, ``Rudimentary quasicrystallography: The icosahedral and decagonal reciprocal lattices," Physical Review B 35.11 (1987): 5487.

\bibitem{Penrose74}  R.~Penrose, ``The role of aesthetics in pure and applied mathematical research," Bull.\ Inst.\ Math.\ Appl.\ {\bf 10}, 266 (1974). 

\bibitem{Gardner77} M.~Gardner, ``Extraordinary nonperiodic tiling that enriches the theory of tiles," Sci.\ Amer.\ {\bf 236}, 110 (1977).

\bibitem{Penrose78} R.~Penrose, ``Pentaplexity," Eureka {\bf 39}, 16 (1978)

\bibitem{BoyleSteinhardt1} L.~Boyle and P.J.~Steinhardt, ``Self-Similar One-Dimensional Quasilattices," arXiv:1608.08220.

\bibitem{SocolarSteinhardt86} J.~E.~S.~Socolar and P.~J.~Steinhardt, ``Quasicrystals. II. Unit-cell configurations," Phys.\ Rev.\ B {\bf 34}, 617 (1986).

\bibitem{Socolar89} J.~E.~S.~Socolar, ``Simple octagonal and dodecagonal quasicrstyals," Phys. Rev. B {\bf 39}, 10519 (1989).

\bibitem{BaakeRootLattices} M.~Baake {\it et al}, ``Root lattices and quasicrystals," J. Phys.\ A: Math.\ Gen. {\bf 23}, L1037 (1990).

\bibitem{ElserSloane} V.~Elser and N.J.A.~Sloane. ``A highly symmetric four-dimensional quasicrystal." Journal of Physics A: Mathematical and General 20.18 (1987): 6161.

\bibitem{SadocMosseri1} J.F.~Sadoc and R.~Mosseri. ``The E8 lattice and quasicrystals: geometry, number theory and quasicrystals." Journal of Physics A: Mathematical and General 26.8 (1993): 1789.

\bibitem{SadocMosseri2} J. F.~Sadoc and R.~Mosseri. ``The E8 lattice and quasicrystals." Journal of non-crystalline solids 153 (1993): 247-252.

\bibitem{MoodyPatera} R.V.~Moody, and J.~Patera. ``Quasicrystals and icosians." Journal of Physics A: Mathematical and General 26.12 (1993): 2829.

\bibitem{BaakeGahler} M.~Baake and F.~Gahler, "Symmetry Structure of the Elser-Sloane Quasicrystal," in Aperiodic 97, eds. M. de Boissieu et al., World Scientific, Singapore (1998), pp. 63-67 [arXiv:cond-mat/9809100].

\bibitem{Reiner} I.~Reiner, {\it Maximal Orders} (Academic Press, New York/London, 1975).

\bibitem{Tits} J.~Tits, ``Quaternions over $\mathbb{Q}(\sqrt{5})$, Leech lattice, and the sporatic group of Hall-Janko, J.~Algebra {\bf 63}, 56 (1980).

\bibitem{Vigneras} M.-F.~Vigneras, {\it Arithmetique des algebres de quaternions}, Lecture Notes in Mathematics, Vol.~800 (Springer-Verlag, New York/Berlin, 1980).

\bibitem{MoodyWeiss} R.V.~Moody and A.~Weiss. ``On shelling E 8 quasicrystals." Journal of Number Theory 47.3 (1994): 405-412.

\bibitem{ConwaySmith} J.H.~Conway and D.A.~Smith, {\it On Quaternions and Octonions: Their Geometry, Arithmetic, and Symmetry} (A K Peters, Natick, MA, 2003). 

\bibitem{WashingtonCyclotomic} L.C.~Washington, {\it Introduction to Cyclotomic Fields} (Springer-Verlag, New York, 1997).

\bibitem{Edwards} H.M.~Edwards, {\it Fermat's Last Theorem} (Springer-Verlag, New York, 1977).

\bibitem{Neukirch} J.~Neukirch, {\it Algebraic Number Theory} (Springer-Verlag, Berlin, 1999).

\bibitem{Humphreys2} J.E.~Humphreys, {\it Introduction to Lie Algebras and Representation Theory} (Springer-Verlag, New York, 1972).

\bibitem{Gabriel} P.~Gabriel, Unzerlegbare darstellungen I., Manuscripta Math.\ {\bf 6}, 71 (1972).

\bibitem{Arnold} V.I.~Arnold, {\it Catastrophe Theory}, 3rd ed., (Springer-Verlag, Berlin, 1992).

\bibitem{ADEintro} M.~Hazewinkel, W.~Hesselink, D.~Siersma, and F.D.~Veldkamp. ``The ubiquity of Coxeter-Dynkin diagrams." Nieuw Archief voor Wiskunde {\bf 25}, 257 (1977).

\end{thebibliography}
\end{document}